# From artificial to circular intelligence to support the well-being of our habitat

Larosa, Francesca[1,2*], Depellegrin, Daniel[3], Conte, Andrea[4], Molinari, Marco[5], Wickberg, Adam[6], Santato, Silvia[7], Mallor, Fermin[1], Sperotto, Anna[8]

[1] FLOW Engineering Mechanics, Royal Institute of Technology (KTH), Stockholm, Sweden
[2] Climate Action Center, Royal Institute of Technology (KTH), Stockholm, Sweden
[3] IHCantabria - Instituto de Hidráulica Ambiental de la Universidad de Cantabria, Santander, Spain
[4] Andreco Studio and Futurecologies srl, Rome, Italy
[5] Department of Energy Technology, Royal Institute of Technology (KTH), Stockholm, Sweden
[6] Centre of excellence for Anthropocene History, KTH Royal Institute of Technology, Stockholm, Sweden
[7] Sea the Change SrL, Rinascimento Green, Italy
[8] Department of Environmental Sciences, Informatics and Statistics (DAIS), Ca' Foscari University of Venice, Venice, Italy

* Corresponding author: larosa@kth.se

**Abstract**
The proliferation of machine learning and artificial intelligence is redefining the interaction between the anthropogenic and natural elements of our habitat. The use of monitoring tools, processing facilities and IoT supports the assessment of our planet's health at any given time through automation. However, these data-, natural resources- and infrastructure-intensive technologies are not neutral on the Earth. As the community of AI practitioners is working on the creation of tools with minimal socio-environmental impact, we contribute to these efforts and we propose a novel conceptual and procedural framework called Circular Intelligence (CIntel). CIntel leverages a bottom-up and community-driven approach to learn from the ability of nature to regenerate and adapt. CIntel incorporates ethical principles in its technical design to preserve the stability of the habitat, while also increasing the well-being of its inhabitants by design. CIntel is operationalized through a set of economic incentives which promote a shared-cost-distributed benefits paradigm.

**Introduction**
The impacts of climate change and biodiversity loss are to a large extent anthropogenic. Humankind has learnt to extract natural resources to support and expand its survival and livelihood, transforming them into novel elements through innovation, industry, research and practice. With progress, over-exploitative practices came to life and the era of the *Great Acceleration*, the Anthropocene[1], quickly escalated towards a hierarchical dominance of humans over biotic and abiotic components of the planet. The development, use and deployment of artificial intelligence (AI) and machine-driven algorithms is an integral part of the trajectory of the Great Acceleration, as fast-growing computational resources facilitate the continued expansion of anthropogenic control over nature[2]. AI and machine learning (ML) help monitoring the fragile equilibrium generated by the constant interplay between anthropogenic and natural elements[3]. However, they also create new stressors on the society and the environment due to their energy and material requirements and their data collection practices. Infrastructure intensity has a direct impact on life below water and on land (e.g. there are 1.8 million km of telecommunication subsea cables connected by 1636 landing stations[4]) algorithms are electricity-thirsty in both the training and inference stages[5]. Concerns about potential rebound effects are also on the rise: more accessible and facilitated consumption and production patterns have unclear consequences on sustainability. AI is also a data-hungry technology[6]: individual and collective information is harnessed at every given time and this leads to potential privacy and copyright violations which must be prevented and accounted for.

Over the past years, a lively debate has shaken the AI policy arena. Regulations, strategies and declarations have flourished[7] to delineate what a safe, ethical and responsible AI is and entails. However, very few implementation protocols has followed through revealing that cooperation will be essential to move from "what shall be regulated" to "how to regulate". New human-machine-nature interactions increase per number, size and complexity across multiple societal domains. The habitat – which we define as the space of co-existence, interaction and integration of these three components – will change and so the risks and opportunities connected to it. We urge a workable and operational shift to expand societal development and progress while also placing the well-being of the habitat at its heart. We call it Circular Intelligence. Circular Intelligence is a principles-based framework which informs AI deployment and has the goal to trigger transformative sustainable processes through the activation of virtuous human-machine feedback loops. Circular intelligence is a bottom-up and community-grounded framework that works to preserve the resilience of the biosphere, protect it from shocks and enhance interspecies well-being while simultaneously accounting for negative impacts of human development.

Circular intelligence is a ***concept***: as it happens in natural systems, circular intelligence mirrors nature's functions to restore, repair, reuse and regenerate different components of the system without producing any waste, while including technological advancements[8]. Circular intelligence is a ***process:*** it aims at governing the interaction within the habitat, through the collection and use of data. However, following circular economy dictates, information generated from concluded actions are re-introduced in the framework. Circular intelligence is a ***common goal:*** to work, it needs intent, which serves as a prompt informed by principles of justice, equity, fairness and environmental protection. These ethical dimensions change overtime and strongly depend on the context they are applied to. Therefore, circular intelligence requires open democratic processes to discuss, review and reshape the framework.

Circular intelligence is multidisciplinary by nature and inherently participatory. It was developed by a diverse group of researchers and scholars working on artificial intelligence, environmental humanities, nature-based solutions, blue and green ecosystem services, public art and collective practices. As a concept, circular intelligence has been sketched through group discussions and dynamic activities over a five-day workshop in Stockholm (May 2024). As a process, circular intelligence has materialised through the artistic representation of the overall framework. Interactions between different elements spanning multiple simultaneous scales and domains have become apparent thanks to the abstraction efforts of a climate-focused artist (Figure 1). As a common goal, circular intelligence represented a compromising solution between often conflicting visions (i.e., technocentrism and ecocentrism), foci (i.e., built and nature-based environment) and approaches (i.e., qualitative vs quantitative).

**The Anthropocene and the mediated planet**

The world just entered the "era of global boiling"[9] with mounting biodiversity losses: it is more urgent than ever to limit as much as possible global warming below 1.5C above pre-industrial levels and to prevent the destruction of ecosystems across the globe. Digitalisation (including AI, IoT technologies and ML) contributes to climate mitigation and adaptation[10,11] and stimulates the expansion of the digital economy, meant as the set of economic and social activities that are enabled by digital technology. Digital technologies reduce physical distances, enabling trade and non-monetary exchanges from the most resilient to highly vulnerable communities.

The birth and expansion of AI and ML opens new opportunities in the data collection, processing and usage stages. The Earth itself is understood and studied thanks to measurement tools which continuously harness information. Sensor networks, IoTs, data buoys and remote surveillance tools are redefining terrestrial and marine habitat protection, species tracking and environmental monitoring. Satellites and remote sensing tools are at the heart of quasi real-time assessments of disruptive events and support conservation efforts wherever needed[12]. For instance, The Ocean Vital Signs Network Initiative promoted by the Woods Hole Oceanographic Institute (https://www.whoi.edu/) is the first large scale Ocean's IoT based on a network of moorings and sensors that provides 4-dimensional observations of the ocean's carbon cycling processes and health. Digital Twins, a real-time digital replica of the earth system integrating both historical and live data will provide necessary information for actionable decision-making and planning. This ongoing transition has become so comprehensive that the knowledge, management and governance of the Earth system became a "mediated planet"[2], dependent on enormous flows of data from a "vast machine" of measuring tools[13]. As part of the habitat, anthropogenic activities are also highly datafied. Distributed sensing networks and AI are radically changing the way humans live in the systems they create and live. Monitoring systems are becoming pervasive across different societal domains, from healthcare[14] to buildings and cities[15]. These tools help create a new understanding of interaction between people themselves and their socio-economic environments. These cyber-physical-human systems ultimately support informed decision making and optimized planning at different levels[16].

The increasing amounts of data being produced and modeled over the past decade constitutes a part of the Great Acceleration: users are also active data suppliers themselves leading to algorithms enriching their knowledge base with detailed and updated databanks. At the same time, AI and ML fed with this information are net beneficiaries of natural resources being power and material-hungry by construction[17]. To reverse this trend, we need to radically rethink the design and purpose of our digital technologies. This is where circular intelligence comes in. Data is solicited according to protocols strongly dependent on the set of interests, skills and processing facilities available at a given point in time and space. As such, data collection becomes a **political process** which feeds into and modifies the data processing routines. AI and ML enable the harmonisation and analysis of dishomogeneous sources, leveraging on the multi-mode power to translate diverse data points into the universal language of mathematics and linear algebra. While this is procedurally true, human intervention is not ruled out as AI and ML-enabled data processing require decisions to be made since the scraping activities. Data processing is also highly shaped by codified ***intents***, meant as the set of motives, beliefs and values embedded in specific communities[18]. Finally, data usage and storage - as the ending point of the value chain - is influenced by socio-techno-cultural forces including market-driven mechanisms and economic motivations of profit, demand and supply. Throughout the whole AI and ML data production chain the spectrum of harmful biases and unethical considerations remains solid and adds to the ballooning environmental concerns surrounding AI development[5].

The non-neutral features of AI on the habitat have been left untackled for quite some time. More fundamentally, the growing trend of AI applied to sustainability challenges is not confronting the issues of an ill-planned machine-augmented society. Circular intelligence challenges this vision and offers an ***ex ante*** reflection on how to design, develop, deploy and decommission machine-driven infrastructures compatible with and integrated intoh the natural ecosystems. Rather than problematising AI once it is applied to real life problems, circular intelligence aims at designing an ethical-rich intelligence for and from the habitat.

**Challenging the established AI framework: the circular intelligence pillars**
The emergence and advancement of AI and ML across multiple sectors of our economy have eroded the boundaries among physical, digital and biological worlds, altering the habitat and changing the interactions between the anthropogenic and natural components. AI has attracted scholarly attention with respect to efficiency gains and automation opportunities across multiple sectors and domains[19]. Since the market launch of ChatGPT and the expansion of generative AI, efficiency and efficacy improvements have been paired to new capability generation with unknown – but potentially disruptive effects – on firms' value propositions. The uptake and development of AI-based tools have followed a relatively uncontrolled scaling trajectory[20] with abrupt consequences from an ethical and environmental point of view. The socio-economic benefits that AI promises are counterbalanced by AI threats, incidents and hazards derived from malfunctioning, biases and malicious uses[21].

The dominant risk management frameworks in AI systems typically focus on the governance aspects across the AI value chain. Issues connected to ethics, responsibility, equity and trustworthiness of the AI technology are at the forefront of initiatives launched by regulatory bodies with legislative purposes (including the EU AI Act[22] and the national policies). They may be promoted by non-governmental organisations (i.e., the Toronto Declaration) or they can be embodied by academic gatherings (i.e., NeurIPS Ethics General Guidelines[23]). These frameworks suggest to ban or restrict AI systems based on the acceptability levels of risks against humans. While being a highly relevant first step in the direction of a collective debate, they fail at providing operational guidelines on how to create more – by design - ethical, responsible, governable, reliable, traceable and trustworthy alternative systems. To address safety and reliability concerns, the AI community has implemented guardrails[24] which maintain alignment between machine-based systems and ethical principles. Guardrails can prevent Large Language Models (LLMs) from processing harmful requests: they act as watchdog algorithms monitoring inputs and outputs. As effectiveness is largely dependent on the AI company or programmer's set of values, the design of guardrails does not follow standard protocols exceptions made for common sense requirements (e.g., hallucination and toxicity)[25]. A more distributed, open and democratic conversation about the power dynamics behind this (mis)alignment is still far from becoming reality[26]: few skilled elites in the West control the data market making many datasets used for training purposes unfit for other contexts[27]. Alternative approaches are presented in the form digital commons[28] (i.e., communal ownership of knowledge and information resources) and include data, compute capacity and skills[29].

The environmental footprint of AI systems is also consistently rising as a major preoccupation[30]. Progress in AI is achieved through scaling deep models and their training data with significant energy[31,32], resources[33] and distributional[34] implications. The electricity needed to power data centers in the United States only is projected to reach 12 percent of total domestic demand by 2028[35]. Using life-cycle assessment methods, research has shown that transitioning from classic digital services to generative AI-based ones would severely increase the environmental footprint of the ICT sector: an open source text-to-image model used for one year depletes metals traceable in 5659 smartphones and emits the equivalent carbon emissions of 48 homes in the US for a year[36]. Water consumption (i.e., "evaporated, transpired, incorporated into products or crops, or otherwise removed from the immediate water environment"[37]) associated to the AI industry amplifies the AI's footprint across the whole supply chain. Water serves as coolant in data centers both in training and inference processes (Scope-1), but it is also a crucial resource in electricity generation (Scope-2). The expansion of AI technologies intensifies competition for water resources, with projections suggesting that the combined Scope-1 and Scope-2 water withdrawal demands of the global AI industry could equate to four to six times the annual water withdrawals of a country the size of Denmark. Scope-3 embodied water consumption is also largely underestimated. The semiconductor industry – for which ultrapure water and water in wafer production is crucial – is classified as having "critical impact" across the whole value chain[38]. In spite of the energy and resource efficiency publicly advertised by the industry[39], the issues connected to the environmental footprint of AI extend way beyond its direct impacts. Efficiency gains can spur increased consumption and production according to the well-known "Jevon's paradox"[34]. Due to its ubiquitous nature, AI's rebound and second-order effects are difficult to assess. From a theoretical point of view, quantitative scenarios and economic projections are not fully incorporating AI within their design; qualitative exercises are also severely lacking plausible pathways of change[40].

Drawing on these ethical and environmental insights as a foundation, we acknowledge that to channel the transformative potential of AI technologies towards a sustainable and responsible use an interdisciplinary and comprehensive new framework is required. Our circular intelligence model, technological, ethical and environmental factors do not operate in siloes, but they critically shape each other. From a technical and technological point of view, dominant models benefit from existing hardware trends without robust underlying motivation[41]. Existing architectures can and shall be improved with environmental parameters becoming active criteria of performance evaluation. Given data, infrastructure and electricity are increasingly scarce commodities[29], the future of AI largely depends on the adaptation potential of systems subject to limited resource availability. Our Circular Intelligence model proposes to incorporate life-cycle and ecosystem services value assessments to design environment-mindful machine systems. While financial disbursement is centralized and incurred where and when training and infrastructure is located, environmental impacts are distributed and diluted overtime. Circular intelligence integrates ethics and a novel system of economic incentives to promote alignment between gains and motives. To counterbalance the temporal mismatches between effective and perceived impacts, circular intelligence benefits from recurring monitoring checks, which help communities resetting their priorities.

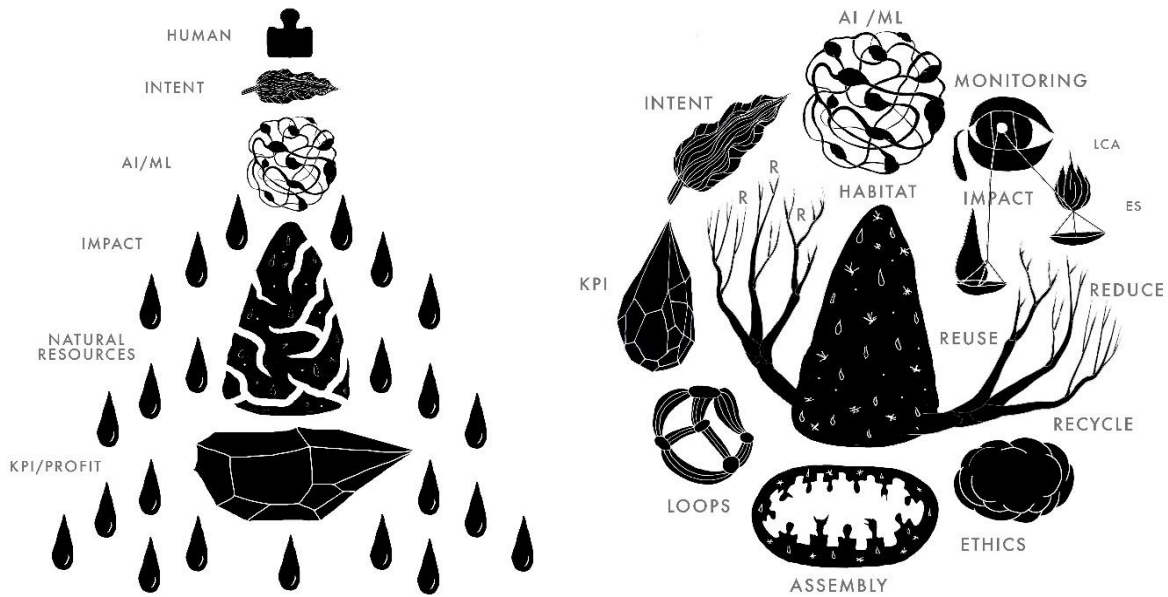

**Figure 1.** Circular Intelligence elicited as artistic practice highlights the role of community, principles and circularity which shape all stages of the AI workflow. Circular Intelligence acknowledges the presence of influence of positive and feedback loop which strongly influence the habitat and the need for monitoring the environmental impacts. In the idea generation process, art has been the instrument to collect inputs and feedback, but also to give birth to a shared understanding of the pillars and functionalities of a new paradigm.

At the heart of modern AI lie neural networks: means through which computers learn and perform specific tasks through the analysis of training databases. In the supervised learning case, training samples are not immediately available in the habitat, but they require human intervention: large-scale labeling efforts inform object recognition and classification tasks, while manual checks correct outputs if needed. Training is a milestone activity of AI systems and applications and the material used to enable it has often been criticized as lacking transparency[43] or expression of partial or biased viewpoints[44]. Efforts to reverse this trend include the publication of training-relevant and task-specific databases (https://www.dataprovenance.org/), but communities may indulge in progressively polarised views due to their use of unfit AI systems[45]. In the reinforcement learning case, agents learn their behaviour by maximising a cumulative reward while minimising penalties in response to wrong or inaccurate actions. In both learning techniques, a more shared and collectively accepted understanding of what a community can perceive as relevant and long-term rewarding can affect AI learning pathways. This is the first novelty of the circular intelligence approach: shifting the collective agency from passive recipient of outputs into active co-creator of AI inputs. Collective engagement opportunities (e.g., in the form of citizens assemblies and living labs), democratise access to technology and redirect digital innovation towards locally-fed AI tools to advance shared long-term goals.

Communities do not just act as decision-makers in the training phase. They also decide over the deployment of AI tools scrutinizing the goals they intend to pursue through intent. Intent responds to ethical considerations which are (as in the previous step) representative of the collective well-being. As modern AI is developed as a technological advancement, some applications lead to unintended negative consequences for both the people and the planet. Circular intelligence internalises risks since its design and mirrors the ability of natural ecosystems to adapt to and regenerate from exogenous shocks by considering the consequences of its deployment. Specifically, circular intelligence accounts for the environmental costs linked to its development and the potential negative loops which may alter the habitat's equilibrium. AI systems require natural resources in their core components (i.e., materials) and throughout their functional life; they are highly energy-intensive both in the training and inference stages[46] and they may fuel overconsumption patterns[40]. At every cycle, circular intelligence computes these costs as relative to the benefits it produces for the habitat. To achieve this goal, circular intelligence integrates methods and approaches from Life Cycle Assessment for material accounting and Ecosystem Services and Nature-based solutions for value quantification. As for nature-based solutions (i.e., solutions that are inspired and supported by nature, which are cost-effective and simultaneously provide environmental, social and economic benefits and help build resilience[47]), circular intelligence runs with an ex-ante estimate of costs discounted overtime to avoid putting in place ex-post actions to repair damaged ecosystems and broken societies. Technically, circular intelligence architecture shall i) **reduce** its negative impacts on the environment by discounting energy and material use costs in its training and inference processes. Whenever emissions are released and resources deployed, circular intelligence shall put in place compensatory and redistributive mechanisms; ii) **re-use** existing datasets, algorithms and methods to continuously update findings. The habitat is in constant evolution due to endogenous and exogenous shocks. Circular intelligence makes use of existing datasets, algorithms and approaches to provide community-relevant insights over a prolonged period of time. To achieve this goal, circular intelligence browses from a wide-scale cloud in which inputs are stored and maintained. This vision calls for a new governance of technology and digitalisation where AI least advanced areas of the world (i.e., the EU) cooperate with the most advanced players (i.e., US-based) to overcome infrastructure challenges; iii) **recycle** functional data and algorithm protocols to build a shared and accepted architecture. The application of the FAIR principles to the AI domain (as suggested by Huerta et al. 2023[48]) is a key enabler of this logic as it moves towards interoperable and modular standard systems. Successful examples already exist: the ARIES project (https://aries.integratedmodelling.org/) is enhancing an open and usable set of interoperable models in environmental sustainability[49], while OASIS Loss Modeling Framework (https://oasislmf.org/) is advancing disaster risk and disaster insurance; iv) **re-visit** circular intelligence's features if community's desiderata change and if the deployment of AI tools alter the habitat. Monitoring and evaluation protocols shall consider whether the use of AI tools shifts the equilibria or lead to unintended shocks. These protocols are also useful in designing effective compensatory mechanisms which account for the redistribution of benefits to ensure equity is in place. Circular intelligence moves beyond existing efforts (i.e, the OECD AI Incidents Monitor database, https://oecd.ai/en/incidents) and includes environmental footprint and impact accounting to secure alignment between intent and AI deployment.

**Making circular intelligence a reality**
Once designed according to circular intelligence's principles (reduce, re-use and re-visit), AI systems become available to the community which shall control and exert its decisional power to pick solely those AI tools which align to the value system represented in a specific group. Circular intelligence borrows from two existing examples, reversing their conceptual framework: data communities and Web3 technologies. Data communities use a bottom-up approach to leverage the power of the crowd in the data collection phase. They are participatory in nature to empower individuals and groups, collect diverse perspectives to be more inclusive, enhance data quality by peer monitoring and provide more control and ownership over the content generated. Well-known and successful data communities have improved knowledge production (i.e., Habitat Map, www.habitatmap.org), accessibility (i.e., Wikipedia) and discoverability (i.e., Zooniverse, https://www.zooniverse.org/about). Web3 technologies enable the diffusion of a new and decentralised internet built on distributed and community-controlled blockchain modules. As in the data communities and Web3 tech cases, circular intelligence puts the community at its heart and empowers individuals and groups through a participatory process aimed at deploying those tools aligned with community values.

Circular intelligence, as built on decentralized systems, call for a *shared costs and distributed benefits* (SCDB) procedural mechanism. Each newly trained AI incurs in environmental and monetary costs which are typically beard by the developer. The community first proposing the design, creation and launch of the AI invests upfront in the training costs, which also account for the environmental footprint of the process. Wealthy and resourceful communities will be the ones capable of innovating, as lower income and vulnerable ones would be in no position to support the initial costs. However, the SCDB mechanism incentives the distribution of the AI systems to as many users as possible to distribute the costs among beneficiaries. In order to build an equitable society, participants to the scheme contribute to the costs unevenly with wealthier agents assuming more of the whole burden. As use cases increase and the technology spreads, beneficiaries take-up part of the economic obligations, while also benefitting from its functionalities. To ensure community values are fully embedded and represented, a bottom-up discussion shall always take place before approving adoption.

The SCDB mechanism fuels an already lively call for a more decentralised[50], ethical[51] and environmental-friendly[52] AI. Differently from existing experiences, circular intelligence proactively proposes a shift in the way AI is developed, owned, controlled and monitored. By doing so, it contributes to a constructive exchange on the type of governance we want for the future to bring communities at the forefront of technological innovation. Circular intelligence holistically welcomes these calls and expands them in a more radical proposal: rather than facing future negative impact with risk-based frameworks, circular intelligence accounts for environmental and social costs since the very design phase of AI systems using a collective and bottom-up approach. Further, circular intelligence does not adhere to a technology-oriented regulatory framework, but poses at the center of the debate the use of each application independently from the AI task or model. Finally, circular intelligence is grounded in the idea of interconnected systems and acknowledges that anthropogenic and natural elements are unequivocally and intrinsically linked in a fragile equilibrium. To make machine-based innovations at use of the well-being of the habitat, multiple scientific domains need to come together bringing their unique perspectives to the discussion. Beyond usual scientific domains, artistic practices are instrumental to communicating and resonating with the communities circular intelligence tries to reach. In our workshop, the inclusion of visual creative elements has led to adequate level of abstraction to find a common ground across different disciplines, jargons and understandings.

**Funding statement**
FL and AS acknowledge funding from the Lerici Foundation Academic Grant 2023-2024. FL acknowledges funding from the European Union's Horizon Europe research and innovation programme under the Marie Skłodowska-Curie grant agreement No. 101150729. AC acknowledges funding support from the Future Environment project, granted by the Italian Council program (2024) promoted by the Directorate-General for Contemporary Creativity of the Italian Ministry of Culture. MM acknowledges funding from the Swedish Energy Agency, project number P2023-01513, and from Digital Futures, project HiSSx, project agreement KTH-RPROJ-0146472. AS acknowledges funding from the European Union Next-GenerationEU in the framework of the iNEST - Interconnected Nord-Est Innovation Ecosystem (iNEST ECS_00000043 – CUP H43C22000540006). DD acknowledges funding support from the Ramón y Cajal grant RYC2022-035260-I, awarded by the Spanish Ministry of Science and Innovation (MCIN/AEI/10.13039/501100011033) and by the European Social Fund Plus (ESF+).

**References**
1. Steffen, W. *et al.* Sustainability. Planetary boundaries: guiding human development on a changing planet. *Science (80-. ).* **347**, 1259855 (2015).
2. Wickberg, A. *et al.* The mediated planet: Datafication and the environmental SDGs. *Environ. Sci. Policy* **153**, 103673 (2024).
3. European Union. Destination Earth EU. https://destination-earth.eu/.
4. Clare, M. A. *et al.* Climate change hotspots and implications for the global subsea telecommunications network. *Earth-Science Rev.* **237**, 104296 (2023).
5. Crawford, K. Generative AI's environmental costs are soaring — and mostly secret. *Nature* (2024) doi:10.2307/j.ctvpj76dk.6.
6. Bates, D. W. *An Artificial History of Natural Intelligence*. (2024).
7. OECD AI. OECD AI Policy Database. (2024).
8. Suárez-Eiroa, B., Fernández, E., Méndez-Martínez, G. & Soto-Oñate, D. Operational principles of circular economy for sustainable development: Linking theory and practice. *J. Clean. Prod.* **214**, 952 – 961 (2019).
9. UN News. Hottest July ever signals 'era of global boiling has arrived' says UN chief. *Global Perspective human stories* (2023).
10. IPCC. *Climate Change 2022: Mitigation of Climate Change. Contribution of Working Group III to the Sixth Assessment Report of the Intergovernmental Panel on Climate Change*. (2022).
11. Leal Filho, W. *et al.* Deploying artificial intelligence for climate change adaptation. *Technol. Forecast. Soc. Change* **180**, 121662 (2022).
12. Almeida, B., David, J., Campos, F. S. & Cabral, P. Satellite-based Machine Learning modelling of Ecosystem Services indicators: A review and meta-analysis. *Appl. Geogr.* **165**, 103249 (2024).
13. Edwards, P. N. *A Vast Machine*. (MIT Press, 2010).
14. Motwani, A., Shukla, P. K. & Pawar, M. Ubiquitous and smart healthcare monitoring frameworks based on machine learning: A comprehensive review. *Artif. Intell. Med.* **134**, 102431 (2022).
15. Peldon, D., Banihashemi, S., LeNguyen, K. & Derrible, S. Navigating urban complexity: The transformative role of digital twins in smart city development. *Sustain. Cities Soc.* **111**, 105583 (2024).
16. Annaswamy, A. M., Johansson, K. H. & Pappas, G. Control for Societal-Scale Challenges: Road Map 2030. *IEEE Control Syst. Mag.* **44**, 30–32 (2024).
17. Lannelongue, L. *et al.* GREENER principles for environmentally sustainable computational science. *Nat. Comput. Sci.* **3**, 514–521 (2023).
18. Trotta, A., Ziosi, M. & Lomonaco, V. The future of ethics in AI: challenges and opportunities. *AI Soc.* **38**, 439–441 (2023).
19. Dwivedi, Y. K. *et al.* Evolution of artificial intelligence research in Technological Forecasting and Social Change: Research topics, trends, and future directions. *Technol. Forecast. Soc. Change* **192**, 122579 (2023).
20. Haefner, N., Parida, V., Gassmann, O. & Wincent, J. Implementing and scaling

artificial intelligence: A review, framework, and research agenda. *Technol. Forecast. Soc. Change* **197**, 122878 (2023).
21. OECD AI Policy Observatory. OECD AI Incidents Monitor (AIM). *https://oecd.ai/en/incidents* (2025).
22. EUR-LEX. *The EU AI Act*. (2024).
23. NeurIPS. NeurIPS Code of Ethics. Guidelines for Reviewers. at https://neurips.cc/Conferences/2023/EthicsGuidelinesForReviewers (2023).
24. Ayyamperumal, S. G. & Ge, L. Current state of LLM Risks and AI Guardrails. *arXiv Comput. Sci.* (2024).
25. Taubenfeld, A., Dover, Y., Reichart, R. & Goldstein, A. Systematic Biases in LLM Simulations of Debates. in *EMNLP 2024* (2024).
26. Larosa, F. & Wickberg, A. Artificial Intelligence can help Loss and Damage only if it is inclusive and accessible. *npj Clim. Action* **3**, 59 (2024).
27. Longpre, S. *et al.* A large-scale audit of dataset licensing and attribution in AI. *Nat. Mach. Intell.* **6**, 975–987 (2024).
28. Birkinbine, B. J. Commons praxis: Towards a critical political economy of the digital commons. *TripleC* **16**, 290–305 (2018).
29. Verdegem, P. Dismantling AI capitalism: the commons as an alternative to the power concentration of Big Tech. *AI Soc.* **39**, 727–737 (2024).
30. COP29. COP29 Declaration on Green Digital Action. (2024).
31. Luccioni, S. *et al.* Light bulbs have energy ratings — so why can't AI chatbots? *Nature* **632**, 736–738 (2024).
32. International Energy Agency. *What the Data Centre and AI Boom Could Mean for the Energy Sector*. https://www.iea.org/commentaries/what-the-data-centre-and-ai-boom-could-mean-for-the-energy-sector (2024).
33. Bashir, N. *et al.* The Climate and Sustainability Implications of Generative AI. *An MIT Explor. Gener. AI* 1–45 (2024) doi:https://doi.org/10.21428/e4baedd9.9070dfe7.
34. Luccioni, S., Strubell, E. & Crawford, K. From efficiency gains to rebound effects: the problem of Jevons' Paradox in AI's polarized environmental debate. *arXiv* **1**, 1–19 (2025).
35. Shebabi, A. *et al.* 2024 United States Data Center Energy Usage Report. *Lawrence Berkeley Natl. Lab.* (2024).
36. Berthelot, A., Caron, E., Jay, M. & Lefèvre, L. Estimating the environmental impact of Generative-AI services using an LCA-based methodology. *Procedia CIRP* **122**, 707–712 (2024).
37. Li, P., Yang, J., Islam, M. A. & Ren, S. Making AI Less 'Thirsty': Uncovering and Addressing the Secret Water Footprint of AI Models. *arXiv Comput. Sci.* 1–16 (2023).
38. CDP. CDP Water Watch. (2025).
39. NVIDIA. *Sustainability Report: Fiscal Year 2024*. https://www.nvidia.com/en-us/sustainability/ (2025).
40. Luers, A. *et al.* Will AI accelerate or delay the race to net-zero emissions? *Nature* **628**, 718–720 (2024).
41. Hooker, S. The hardware lottery. *Commun. ACM* **64**, 58–65 (2021).
42. McCarthy, J., Minsky, M. L., Rochester, N. & Shannon, C. E. A Proposal for the Dartmouth Summer Research Project on Artificial Intelligence, August 31, 1955. *AI Mag.* **27**, 12 (2006).
43. Fehr, J., Citro, B., Malpani, R., Lippert, C. & Madai, V. I. A trustworthy AI reality-check: the lack of transparency of artificial intelligence products in healthcare. *Front. Digit. Heal.* **6**, (2024).
44. P. S. , D. V. How can we manage biases in artificial intelligence systems – A systematic literature review. *Int. J. Inf. Manag. Data Insights* **3**, 100165 (2023).
45. Rodilosso, E. Filter Bubbles and the Unfeeling: How AI for Social Media Can Foster Extremism and Polarization. *Philos. Technol.* **37**, 71 (2024).
46. Chauhan, D., Bahad, P. & Jain, J. K. Sustainable AI Environmental Implications, Challenges, and Opportunities. *Explain. AI Sustain. Dev. Trends Appl.* **2**, 1–15 (2024).

47. European Environment Agency. *Nature-Based Solutions in Europe: Policy, Knowledge and Practice for Climate Change Adaptation and Disaster Risk Reduction*. *EEA Report* (2021).
48. Huerta, E. A. *et al.* FAIR for AI: An interdisciplinary and international community building perspective. *Sci. Data* **10**, 487 (2023).
49. Balbi, S. *et al.* The global environmental agenda urgently needs a semantic web of knowledge. *Environ. Evid.* **11**, 5 (2022).
50. Institute, D. *DAIR Research Philosophy*. vol. 1 (2022).
51. Floridi, L. *The Ethics of Artificial Intelligence: Principles, Challenges, and Opportunities*. (2023).
52. Crawford, K. *Atlas of AI*. (Yale University Press, 2022).

**Acknowledgements**

We thanks all the participants to the workshop "Towards a RE-GENERATIVE AI: Charting Artificial Intelligence and Nature-based solutions for climate adaptation strategies" (https://www.kth.se/philhist/historia/omoss/events/kalender/ehl/towards-a-re-generative-ai-charting-artificial-intelligence-and-nature-based-solutions-for-climate-adaptation-strategies-1.1336237) funded by the Lerici Foundation and included in the Marie Skłodowska-Curie grant agreement No. 101150729.

**Conflicts of interest**

We confirm that this work is original and has not been published elsewhere, nor is it currently under consideration for publication elsewhere. To the best of the authors' knowledge, no conflict of interest, financial or other, exists.